\documentstyle[11pt]{article}
\textwidth\def\kon#1#2{\vbox{\halign{##&&##\cr\lower4pt
\hbox{$\scriptscriptstyle\vert$}\hrulefill &\hrulefill\lower4pt
\hbox{$\scriptscriptstyle\vert$}\cr $#1$&$#2$\cr}}}

\title{QUANTIZED HERMITIAN SUPERFIELDS}
\author{Florin Constantinescu \\Fachbereich Mathematik,
\\ Johann Wolfgang Goethe Universit\"at Frankfurt,
\\ Robert-Mayer-Str. 10, D-60054 Frankfurt am Main, Germany
\and Markus Gut and G\"unter Scharf 
\\ Institut f\"ur Theoretische Physik, \\ Universit\"at Z\"urich, 
\\ Winterthurerstr. 190 , CH-8057 Z\"urich, Switzerland}\date{}

\begin{document}
\maketitle\vskip 3cm

\begin{abstract}  We analyse the algebras generated by free component
quantum fields together with the susy generators $Q,\bar Q$. Restricting
to hermitian fields we first construct the scalar field algebra from
which various scalar superfields can be obtained by exponentiation. Then
we study the vector algebra and use it to construct the vector
superfield. Surprisingly enough, the result is totally different from
the vector multiplet in the literature. It contains two hermitian
four-vector components instead of one and a spin-3/2 field similar to
the gravitino in supergravity.

\end{abstract}

\newpage
\def\tr{{\rm tr}}
\def\ro{\varrho}
\def\eh{{\scriptstyle{1\over 2}}}
\def\d{\partial}
\def\=d{\,{\buildrel\rm def\over =}\,}
\def\dh{\mathop{\vphantom{\odot}\hbox{$\partial$}}}
\def\dl{\dh^\leftrightarrow}
\def\sqr#1#2{{\vcenter{\vbox{\hrule height.#2pt\hbox{\vrule width.
#2pt height#1pt \kern#1pt \vrule width.#2pt}\hrule height.#2pt}}}}
\def\w{\mathchoice\sqr45\sqr45\sqr{2.1}3\sqr{1.5}3\,} 
\def\eps{\varepsilon}
\def\oe{\overline{\rm e}}
\def\onu{\overline{\nu}}\def\ds{\hbox{\rlap/$\partial$}}
\def\psq{{\overline{\psi}}}
\def\la{{(\lambda)}}\def\lap{\bigtriangleup\,}
\def\ev{{\scriptstyle{1\over 4}}}
\def\sq{\hbox{\rlap{$\sqcap$}$\sqcup$}}

\section{Introduction}

In this paper we continue our study of quantized free superfields
started in [1]. In paper [1] we have constructed the quantized chiral
superfield and we have found that it has the same component expansion as
its classical counterpart [2-5]. Surprisingly enough, if we go on to
more complicated superfields, this is no longer the case !

To understand this interesting fact we can give the following
explanations. First of all the infinitesimal susy transformation of a
classical superfield $\Phi(x,\theta,\bar\theta)$ is defined by applying
the operator $(\xi Q+\bar\xi\bar Q)$ where $Q,\bar Q$ are represented by
the usual superspace differential operators. On the other hand, the
corresponding transformation of the quantum fields is given by a
commutator
$$\delta\Phi=[\xi Q+\bar\xi\bar Q,\Phi]\eqno(1.1)$$
where now $Q,\bar Q$ are operators in Fock space. Secondly, the
components of the quantum superfield satisfy (anti-)commutation
relations. There is an interesting interplay between these relations,
the free field equations and the supersymmetric algebra, not existing in
the classical case. In fact, the construction of these supersymmetric
free field algebras is the main work to be done. First this is carried
out for the hermitian scalar superfield in the next section and then for
the vector superfield in sect.3. The superfields themselves are obtained
from the algebra simply by exponentiation.
The vector algebra will be used to construct the vector superfield.
The latter comes out totally different from the vector field in the
literature [2-3]. For example, it contains a spin-3/2 component similar
to the gravitino in supergravity and two hermitian four-vector
components instead of one. We are obviously driven to another
representation of the supersymmetric algebra, when we consider quantized
free fields.

\section{The quantized hermitian scalar superfield}

We want to represent the supersymmetric algebra
$$\{Q_a,\bar Q_{\bar b}\}=2\sigma^\mu_{a\bar b}P_\mu=-2i\sigma^\mu_{a\bar
b}\d_\mu,\eqno(2.1)$$
$$\{Q_a, Q_b\}=0=\{\bar Q_a,\bar Q_{\bar b}\}=[Q_a, P_\mu]\eqno(2.2)$$
by operators in Fock space. $Q_a$ and $\bar Q_{\bar b}$ transform 
according to the $(\eh, 0)$ and $(0, \eh)$ representations of the proper
Lorentz group, respectively. As usual one rewrites (2.1) as a Lie-algebra
commutator
$$[\theta^a Q_a,\bar\theta^{\bar b}\bar Q_{\bar b}]=2\theta\sigma 
\bar\theta P=-[\theta Q,\bar\theta\bar Q]\eqno(2.3)$$
by introducing anti-commuting C-numbers $\theta^a, \bar\theta^{\bar b}$.
It is our aim to construct a superfield $S$ with the transformation law 
$$\delta S={\d\over\d\xi}S_\xi\Bigl\vert_{\xi=1}=i[\theta Q+ 
\bar\theta\bar Q,S]\eqno(2.4)$$ 
under infinitesimal supersymmetric transformations. Taking as initial 
value the scalar component field $C(x)$, the solution is given by the 
finite transformation 
$$S_\xi(x,\theta,\bar\theta)=e^{i\xi(\theta Q+\bar\theta\bar Q)}C(x) 
e^{-i\xi(\theta Q+\bar\theta\bar Q)}.\eqno(2.5)$$ 
The factor $i$ has been inserted to get a hermitian field 
$$S_1(x,\theta,\bar\theta)^+=S_1(x,\theta,\bar\theta)\eqno(2.6)$$ 
by assuming $C(x)$ to be hermitian. Then $C(x)$ must be quantized as a 
neutral scalar field of mass $m$ 
$$[C(x), C(x')]=-iD(x-x'),\eqno(2.7)$$ 
where $D(x)$ is the causal Jordan-Pauli distribution for mass $m$. 

We want to calculate (2.5) by means of the Lie series. For the first order
terms we need the commutators 
$$[Q_a,C(x)]=\chi_a(x),\quad [\bar Q_{\bar b},C(x)]=-\bar\chi_{\bar
b}(x), \eqno(2.8)$$ 
where $\chi$ and $\bar\chi$ are quantized Majorana fields. The second 
commutation relation follows from the first by taking the adjoint, 
because (see [1] eq.(2.21))
$$Q_a^+=\bar Q_{\bar a}\eqno(2.9)$$ 
and similarly for $\chi_a$. Next we need the anticommutator 
$$\{ Q_a,\chi_b(x)\}=\{ Q_a,[Q_b,C]\}=\{ Q_b,[C,Q_a]\}-[C,\{ Q_a,Q_b\}] 
=-\{Q_b,\chi_a(x)\},\eqno(2.10)$$ 
where the Jacobi identity has been used. Due to the antisymmetry in $a,b$,
the anticommutator must be of the form 
$$\{ Q_a,\chi_b(x)\}=-im\eps_{ab}F(x),\eqno(2.11)$$ 
where $F(x)$ is a scalar field and the mass factor has been introduced for
dimensional reasons. Since the $\eps$-tensor is real, the adjoint 
relation is 
$$\{\bar Q_{\bar a},\bar\chi_{\bar b}(x)\}=im\eps_{\bar a\bar b}F(x)^+. 
\eqno(2.12)$$ 

The Majorana field $\chi_a$ must be quantized according to 
$$\{\chi_a(x),\chi_b(x')\}=im\eps_{ab}D(x-x'),$$ 
$$\{\chi_a(x),\bar\chi_{\bar b}(x')\}=-\sigma^\mu_{a\bar b}\d_\mu 
D(x-x').\eqno(2.13)$$ 
On the other hand, again by the Jacobi identity the first anticommutator
is equal to 
$$=\{\chi_a(x),[Q_b,C(x')]\}=\{Q_b,[C(x'),\chi_a(x)]\}-[C,\{\chi_a, 
Q_b\}]=$$ 
$$=\{Q_b,[C(x'),\chi_a(x)]\}+im\eps_{ba}[C(x'),F(x)].$$ 
The commutator in the first term must be assumed to be 0, hence we find 
$$[F(x),C(x')]=D(x-x').$$ 
We can therefore write 
$$F(x)=M(x)+iC(x)\eqno(2.14)$$ 
with 
$$[M(x),C(x')]=0.\eqno(2.15)$$ 

We specialize (2.11) to $a=2, b=1$. Using $\eps_{21}=1$ we get 
$$\{Q_2,\chi_1\}=-imF.\eqno(2.16)$$ 
This implies$$[Q_a,F(x)]={i\over m}[Q_a,\{Q_2,\chi_1(x)\}]=$$ 
$$=-{i\over m}([Q_2,\{\chi_1,Q_a\}]+[\chi_1,\{Q_a,Q_2\}])=-\eps_{a1} 
[Q_2,F(x)].\eqno(2.17)$$ 
For $a=1$ this gives 
$$[Q_1,F(x)]=0$$ 
and for $a=2$ 
$$[Q_2,F(x)]=-[Q_2,F(x)]=0.\eqno(2.18)$$ 
The adjoint relations are
$$[\bar Q_{\bar a},F^+(x)]=0.\eqno(2.19)$$

Next we consider 
$$[F(x),F(x')]={i\over m}[F(x),\{Q_2,\chi_1(x')\}]=$$ 
$$=-{i\over m}\{Q_2,[\chi_1(x'),F(x)]\}=0=[M(x),M(x')]-[C(x),C(x')].$$ 
Since the first commutator is $-iD(x-x')$ it follows 
$$[M(x),M(x')]=-iD(x-x'),\eqno(2.20)$$ 
so that $M(x)$ is also a hermitian scalar field. This can also be 
derived from (2.13). To determine the other (anti)commutators we raise 
the second spinor index in (2.12) with the $\eps$-tensor 
$$\{\bar Q_{\bar a},\bar\chi^{\bar c}\}=-im\delta_{\bar a}^{\bar c}F^+. 
\eqno(2.21)$$ 
Taking the trace $\bar a=\bar c=1,2$, we have 
$$F^+={i\over 2m}\{\bar Q_{\bar a},\bar\chi^{\bar a}\}.\eqno(2.22)$$
Let us apply the Dirac operator $i\sigma^\mu_{a\bar c}\d_\mu$ to (2.21) and
use the Dirac equation
$$\{\bar Q_{\bar a},i\sigma^\mu_{a\bar c}\d_\mu\bar\chi^{\bar c}\}=m\{\bar 
Q_{\bar a},\chi_a\}=m\sigma^\mu_{a\bar a}\d_\mu F^+.$$
This leads to
$$\{\bar Q_{\bar a},\chi_b\}=\sigma^\mu_{b\bar a}\d_\mu F^+\eqno(2.23)$$
and the adjoint relation
$$\{Q_a,\bar\chi_{\bar b}\}=\sigma^\mu_{a\bar b}\d_\mu F.\eqno(2.24)$$ 

By means of (2.22) and (2.24) we calculate
$$[Q_a,F^+(x)]={i\over 2m}[Q_a,\{\bar Q_{\bar a},\bar\chi^{\bar a}\}]=$$
$$=-{i\over 2m}\Bigl([\bar Q_{\bar a},\{\bar\chi^{\bar a},Q_a\}]+ 
[\bar\chi^{\bar a},\{Q_a,\bar Q_{\bar a}\}]\Bigl)$$ 
$$={i\over 2m}\sigma^\mu_{a\bar b}[\bar Q^{\bar b},\d_\mu F]+{1\over m} 
\sigma^\mu_{a\bar a}\d_\mu\bar\chi^{\bar a}.\eqno(2.25)$$ 
The adjoint equation is 
$$[\bar Q_{\bar b},F(x)]=-{i\over 2m}\sigma^\mu_{b\bar b}[Q^b,\d_\mu F^+]-
{1\over m}\sigma^\mu_{a\bar b}\d_\mu\chi^a.\eqno(2.26)$$ 
Substituting this into (2.25) and using the relations 
$$\hat\sigma^{\mu\bar aa}=\eps^{\bar a\bar b}\eps^{ab}\sigma^\mu_{b\bar 
b}\eqno(2.27)$$ 
$$(\hat\sigma^\nu\sigma^\mu+\hat\sigma^\mu\sigma^\nu)^{\bar a}_{\bar b} 
=2\eta^{\nu\mu}\delta^{\bar a}_{\bar b},\eqno(2.28)$$ 
we conclude 
$${3\over 4}[\bar Q_{\bar b},F]=-{i\over 2}\bar\chi_{\bar b}-{1\over m} 
\sigma^\mu_{a\bar b}\d_\mu\chi^a.\eqno(2.29)$$ 
In the second term on the r.h.s. we can use the (adjoint) Dirac
equation 
$$\sigma^\mu_{a\bar b}\d_\mu\chi^a=im\bar\chi_{\bar b}.\eqno(2.30)$$ 
This finally gives 
$$[\bar Q_{\bar b},F(x)]=-2i\bar\chi_{\bar b}(x),\eqno(2.31)$$ 
and the adjoint relation 
$$[Q_b,F^+]=-2i\chi_b.\eqno(2.32)$$

Now we are ready to evaluate (2.5) because all multiple commutators on
the r.h.s. are known. The result is
$$S_1(x,\theta,\bar\theta)=C(x)+i(\theta\chi-\bar\theta\bar\chi)+$$
$$-i{m\over 2}(\theta\theta F-\bar\theta\bar\theta
F^+)+\theta\sigma^\mu\bar\theta\d_\mu M+$$
$$+i{m\over 3}(\theta\theta\bar\theta\bar\chi-\theta\chi\bar\theta
\bar\theta)+{1\over 3}\theta\sigma^\mu\bar\theta(\bar\theta\d_\mu
\bar\chi-\theta\d_\mu\chi)+$$
$$+{m^2\over 4}\theta\theta\bar\theta\bar\theta C(x).\eqno(2.33)$$
The third order terms can be rewritten by means of the Dirac equation
and the identity
$$\bar\theta^{\bar a}\bar\theta^{\bar b}=\eh\eps^{\bar a\bar
b}\bar\theta\bar\theta\eqno(2.34)$$
in the form
$$-i{m\over 2}(\bar\theta\bar\theta\theta\chi-\theta\theta\bar\theta
\bar\chi).\eqno(2.35)$$

Another hermitian scalar superfield $S_0$ is obtained as the sum of the 
chiral plus the anti-chiral superfield $\Phi+\Phi^+$ constructed in [1].
This field has a component expansion of the same form as $S_1$ (2.33), 
but is actually different as can be seen by comparing the coefficients. 
The difference is not surprising if one realizes the different behaviour
under infinitesimal susy-transformation. Indeed, from
$$\Phi_\xi=e^{\xi(\theta Q+\bar\theta\bar Q)}A
e^{-\xi(\theta Q+\bar\theta\bar Q)}\eqno(2.36)$$
we find
$$\delta_\xi(\Phi+\Phi^+)={\d\over\d\xi}(\Phi_\xi+\Phi^+_\xi)\Bigl\vert
_{\xi=1}=[\theta Q+\bar\theta\bar Q,\Phi]-[\theta Q+\bar\theta\bar Q,\Phi^+]
,\eqno(2.37)$$
in contrast to (2.4). A third
scalar field $S_2$ can be constructed as $S_1$ (2.5) but starting with the
spinor component
$$S_2(x,\theta,\bar\theta)=e^{i(\theta Q+\bar\theta\bar Q)}(i\theta\chi(x) 
-i\bar\theta\bar\chi(x))e^{-i(\theta Q+\bar\theta\bar Q)}.\eqno(2.38)$$
All commutators are the same as before so that we immediately find
$$S_2=i(\theta\chi-\bar\theta\bar\chi)-im(\theta\theta F-\bar\theta
\bar\theta F^+)
+(\theta\sigma^\mu\bar\theta)(\d_\mu F+\d_\mu F^+)+$$
$$+{3\over 2}im(\theta\chi\bar\theta\bar\theta-\theta\theta\bar\theta
\bar\chi)+{i\over 2}m^2\theta\theta\bar\theta\bar\theta(F^+-F).\eqno(2.39)$$
This scalar field is the most important one because we shall need it when we
consider gauge transformations.
 
Still this is not the whole story because supersymmetry gives further
constraints on the component fields. To see this we take the
commutator of (2.24) with $\bar Q_{\bar c}$ and use (2.31)
$$\sigma^\mu_{a\bar b}[\bar Q_{\bar c},\d_\mu F]=-2i\sigma^\mu_{a\bar b}
\d_\mu\bar\chi_{\bar c}=[\bar Q_{\bar c},\{ Q_a,\bar\chi_{\bar b}\}]=$$
$$=-[Q_a,\{\bar\chi_{\bar b},\bar Q_{\bar c}\}]-[\bar\chi_{\bar b},
\{\bar Q_{\bar c},Q_a\}].$$
Using the relations (2.12),(2.1) and (2.32) the last two commutators can 
be evaluated. This leads to the following relation
$$\sigma^\mu_{a\bar b}\d_\mu\bar\chi_{\bar c}-
\sigma^\mu_{a\bar c}\d_\mu\bar\chi_{\bar b}=-im\eps_{\bar c\bar b}
\chi_a.\eqno(2.40)$$
Here we must only check the nontrivial case $\bar b=1, \bar c=2,$ say.
Then the relation is satisfied due to the Dirac equation. The adjoint 
relation is
$$\sigma^\mu_{b\bar a}\d_\mu\chi_c-
\sigma^\mu_{c\bar a}\d_\mu\chi_b=im\eps_{cb}
\bar\chi_{\bar a}.\eqno(2.41)$$

\section{Quantized vector superfield}

We have learnt in the last section that we do not obtain a vector
component if we start from a scalar field. Therefore, we now start from
a Majorana field $\lambda_a(x)$ as in $S_2$ (2.36), but we do not assume
that $\lambda(x)$ is obtained from a scalar field like (2.8). 
We want to construct the hermitian superfield
$$V(x,\theta,\bar\theta)=e^{i(\theta Q+\bar\theta\bar Q)}(i\theta^a 
\lambda_a(x)+{\rm h.c.})e^{-i(\theta Q+\bar\theta\bar Q)}.\eqno(3.1)$$
If we omit the $\theta^a$ starting from $i\lambda_a$, we obtain a spinor
field $W_a(x,\theta,\bar\theta)$.
To determine the necessary commutators we start with the 
mixed anticommutator $\{Q_a,\bar\lambda_{\bar b}(x)\}$. It must be
proportional to $\sigma^\mu_{a\bar b}$ because there is a one-to-one
correspondence between spinors of type $(\eh, \eh)$ and four-vectors.
Therefore, we introduce two self-conjugate vector fields $v, w$ by requiring
$$\{Q_a,\bar\lambda_{\bar b}(x)\}=-m\sigma^\mu_{a\bar b}(v_\mu 
+iw_\mu).\eqno(3.2)$$
The adjoint relation is
$$\{\bar Q_{\bar a},\lambda_b(x)\}=-m\sigma^\mu_{b\bar a}(v_\mu 
-iw_\mu).\eqno(3.3)$$
As in the last section, two other relations follow by means of the Dirac
equation. We multiply (3.2) by $\eps^{\bar c\bar b}$ and apply the Dirac
operator $i\sigma^\nu_{b\bar c}\d_\nu$. This gives
$$\{Q_a,i\sigma^\nu_{b\bar c}\d_\nu\bar\lambda^{\bar c}\}=m\{Q_a,\lambda
_b\}=-im\sigma^\mu_{a\bar b}\eps^{\bar c\bar b}\sigma^\nu_{b\bar c}
\d_\nu (v_\mu+iw_\mu)=$$
$$=-im\sigma^\nu_{b\bar c}\hat\sigma^{\mu\bar c c}\eps_{ac}\d_\nu
(v_\mu+iw_\mu),$$
or
$$\{Q_a,\lambda_b\}=-i(\sigma^\nu\hat\sigma^\mu)_{ba}\d_\nu
(v_\mu+iw_\mu).\eqno(3.4)$$
The adjoint relation reads
$$\{\bar Q_{\bar a},\bar\lambda_{\bar b}\}=-i(\hat\sigma^\nu\sigma^\mu) 
_{\bar b\bar a}\d_\nu(v_\mu-iw_\mu).$$

Next we have to determine the commutators of $v_\mu$ and $w_\mu$. Taking the
commutator of (3.2) with $Q_b$ we get
$$[Q_b,\{Q_a,\bar\lambda_{\bar b}\}]=-m\sigma^\mu_{a\bar b}
[Q_b,v_\mu+iw_\mu].\eqno(3.5)$$
Since the l.h.s. is antisymmetric in $a,b$ due to the Jacobi identity,
we conclude
$$m\sigma^\mu_{a\bar b}[Q_b,v_\mu+iw_\mu]=-{1\over 2}\eps_{ab}\bar\Lambda 
_{\bar b}\eqno(3.6)$$ 
where
$$\bar\Lambda_{\bar b}=[Q_c,\{Q^c,\bar\lambda_{\bar b}\}].\eqno(3.7)$$
It is easy to see that $\{Q_a,\bar\Lambda_{\bar b}\}=0$ follows and also
the adjoint relation $\{\bar Q_{\bar a},\Lambda_{b}\}=0$. Since
$\Lambda$ satisfies the Dirac equation as $\lambda$, we also have 
$\{Q_a,\Lambda_b\}=0$. Using the Jacobi identity again
$$[\bar Q_{\bar a},\{Q_a,\Lambda_b\}]=0=-[Q_a,\{\Lambda_b,\bar Q_
{\bar a}\}]-[\Lambda_b,\{\bar Q_{\bar a},Q_a\}]=-2i\sigma^\mu
_{a\bar a}\d_\mu\Lambda_b,\eqno(3.8)$$
we finally conclude $\Lambda=0$. Hence, by (3.6)
$$[Q_b,v_\mu+iw_\mu]=0=[\bar Q_{\bar b},v_\mu-iw_\mu],\eqno(3.9)$$
or
$$[Q_b,v_\mu]=-i[Q_b,w_\mu],\quad [\bar Q_{\bar b},v_\mu]= 
i[\bar Q_{\bar b},w_\mu].\eqno(3.10)$$
This corresponds to the result (2.18-19) in the scalar case.

Let us now consider the commutators
$$[Q_a, v^\mu]=-i[Q_a, w^\mu]\=d f^\mu_a,\eqno(3.11)$$
$$[\bar Q_{\bar a}, v^\mu]=i[\bar Q_{\bar a}, w^\mu]\=d -\bar f^\mu
_{\bar a}.\eqno(3.12)$$
From (3.3) we get
$$[Q_c,\{\bar Q_{\bar a},\lambda_b(x)\}]= 
-m\sigma^\mu_{b\bar a}[Q_c, v_\mu(x)-iw_\mu(x)]=-2m\sigma^\mu_{b\bar a}
f_{\mu c}.$$
By the Jacobi identity this is equal to
$$=-[\bar Q_{\bar a},\{\lambda_b,Q_c\}]-[\lambda_b,\{ Q_c,\bar Q_{\bar
a}\}].$$
Using (3.4) and (2.1) we obtain
$$=i(\sigma^\nu\hat\sigma^\mu)_{bc}[\bar Q_{\bar a},\d_\nu
(v_\mu+iw_\mu)]-2i\sigma^\mu_{c\bar a}\d_\mu\lambda_b.$$
This gives an inhomogeneous linear field equation 
$$-i(\sigma^\nu\hat\sigma^\mu)_{bc}\d_\nu\bar f_{\mu\bar a}+ 
m\sigma^\mu_{b\bar a}f_{\mu c}=i\sigma^\mu_{c\bar a}\d_\mu\lambda_b. 
\eqno(3.13)$$
The adjoint equation reads
$$-i(\hat\sigma^\nu\sigma^\mu)_{\bar b\bar c}\d_\nu f_{\mu a}+ 
m\sigma^\mu_{a\bar b}\bar f_{\mu\bar c}=-i\sigma^\mu_{a\bar c}\d_\mu 
\bar\lambda_{\bar b}.\eqno(3.14)$$

With the same technique we treat 
$$\{Q_a,[\bar Q_{\bar b}, v^\mu+iw^\mu]\}=-2\{Q_a,\bar f^\mu_{\bar b}\}$$
$$=-[v^\mu+iw^\mu,\{Q_a,\bar Q_{\bar b}\}]+\{\bar Q_{\bar b},[v^\mu 
+iw^\mu, Q_a]\}$$
$$=-2i\sigma^\nu_{a\bar b}\d_\nu (v^\mu +iw^\mu ).$$ 
This gives the anticommutator
$$\{Q_a,\bar f_{\bar b}^\mu\}=i\sigma^\nu_{a\bar b}\d_\nu 
(v^\mu+iw^\mu),\eqno(3.15)$$
and the adjoint relation
$$\{\bar Q_{\bar a},f_b^\mu\}=-i\sigma^\nu_{b\bar a}\d_\nu 
(v^\mu-iw^\mu).$$

To determine the anticommutator of $f$ we use the differential equation
(3.13). We multiply it by $\hat\sigma^{\ro\bar a a}$:
$$-i\hat\sigma^{\ro\bar a a}(\sigma^\nu\hat\sigma^\mu)_{bc}\d_\nu
\bar f_{\mu\bar a}+m(\sigma^\mu\hat\sigma^\ro)_b^{\, a}f_{\mu c}=
i(\sigma^\mu\hat\sigma^\ro)_c^{\, a}\d_\mu\lambda_b.$$
We put $a=b$ and sum over $b$. Using the trace $\tr (\sigma^\mu
\hat\sigma^\ro)=2\eta^{\mu\ro}$, we obtain
$$-i(\hat\sigma^\ro\sigma^\nu\hat\sigma^\mu)^{\bar a}_{\,c} 
\d_\nu\bar f_{\mu\bar a}+2m f^\ro_c=i(\sigma^\mu\hat\sigma^\ro)_c^{\, b} 
\d_\mu\lambda_b.\eqno(3.16)$$
From (3.14) we find in the same way
$$-i(\sigma^\ro\hat\sigma^\nu\sigma^\mu)_{a\bar c} 
\d_\nu f_\mu^a+2m\bar f^\ro_{\bar c}=-i(\hat\sigma^\ro\sigma^\mu)^ 
{\bar b}_{\,\bar c}\d_\mu\bar\lambda_{\bar b}.\eqno(3.17)$$
The two equations (3.16-17) are similar to the Dirac equation, but
contain three $\sigma$-matrices instead of one. For brevity we will call
the system (3.16-17) the $3\sigma$-equations.

Now we operate with the $3\sigma$-operator $-i(\hat\sigma^\ro\sigma^\alpha
\hat\sigma^\mu)^{\bar b}_{\,c}\d_\alpha$ on (3.15) and substitute (3.16):
$$-2m\{Q_a, f^\ro_c\}+i(\sigma^\mu\hat\sigma^\ro)_c^{\,b}\{Q_a,\d_\mu
\lambda_b\}=$$
$$=\sigma^\nu_{a\bar b}(\hat\sigma^\ro\sigma^\alpha\hat\sigma^\mu)^{\bar
b}_{\,c} \d_\alpha\d_\nu(v_\mu+iw_\mu).$$
Using (3.4) we get
$$-2m\{Q_a,f^\ro_c\}=-(\sigma^\mu\hat\sigma^\ro)_c^{\,b} 
(\sigma^\nu\hat\sigma^\alpha)_{ba}\d_\mu\d_\nu(v_\alpha+iw_\alpha)$$
$$+(\sigma^\nu\hat\sigma^\ro\sigma^\alpha\hat\sigma^\mu)_{ac}\d_\nu 
\d_\alpha(v_\mu+iw_\mu).\eqno(3.18)$$
The result is antisymmetric in $a, c$
$$2m\{Q_a,f^\ro_c\}=\Bigl[(\sigma^\mu\hat\sigma^\ro\sigma^\nu 
\hat\sigma^\alpha)_{ca}-$$
$$-(\sigma^\mu\hat\sigma^\ro\sigma^\nu\hat\sigma^\alpha)_{ac} 
\Bigl]\d_\mu\d_\nu(v_\alpha+iw_\alpha).\eqno(3.19)$$
This antisymmetry is essential for consistency of the algebra because
it follows from $\{Q_a, [Q_c, v^\mu]\}$ by means of the Jacobi identity.

It is not hard to simplify the product of four $\sigma$-matrices. Only
the antisymmetric terms $\sim\eps_{ac}$ survive so
that we finally obtain
$$m\{Q_a,f^\ro_c\}=\eps_{ac}\Bigl[-\sq (v^\ro+iw^\ro)+2\d^\ro\d_\mu
(v^\mu+iw^\mu)\Bigl].\eqno(3.20)$$
It is clear that the second derivatives on the r.h.s. must be simplified
by means of the wave equations for the vector fields. As in quantum
gauge theories [7] we assume the massive vector
fields $v, w$ to satisfy the Klein-Gordon equation
$$\sq v_\nu=-m^2v_\nu,\eqno(3.21)$$
and the same is assumed for $w$.
Then we arrive at the  result
$$\{Q_a, f^\mu_b\}=\eps_{ab}\Bigl[m(v^\mu+iw^\mu)+{2\over m}\d^\mu 
\d_\nu(v^\nu+iw^\nu)\Bigl].\eqno(3.22)$$

To exhaust all consequences of supersymmetry we have to analyse the
commutator
$$[Q_c,\{\bar Q_{\bar a},f^\mu_b\}]=-2i\sigma^\nu_{b\bar a}\d_\nu
f^\mu_c.\eqno(3.23)$$
Using (3.22), (2.1), and the Jacobi identity we find another equation
for $f$
$$i(\sigma^\nu_{c\bar a}\d_\nu f^\mu_b-\sigma^\nu_{b\bar a}\d_\nu
f^\mu_c)=
m\eps_{cb}\bar f^\mu_{\bar a}+{2\over m}\eps_{cb}\d^\mu\d_\nu
\bar f^\nu_{\bar a}.\eqno(3.24)$$
This equation has a similar form as (2.41). Therefore, we simplify it in
the same way as we have treated (2.40). Due to the antisymmetry in $b,
c$, it is sufficient to consider the case $b=1, c=2$. Then the equation
assumes the form
$$i(\sigma^\nu_{2\bar a}\d_\nu f^\mu_1-\sigma^\nu_{1\bar a}\d_\nu f^\mu_2
)=m\bar f^\mu_{\bar a}+{2\over m}\d^\mu\d_\nu\bar f^\nu_{\bar a}.
\eqno(3.25)$$
This is a modified Dirac equation
$$-i\sigma^\nu_{a\bar a}\d_\nu f^{\mu a}=m\bar f^\mu_{\bar a}+
{2\over m}\d^\mu\d_\nu\bar f^\nu_{\bar a}.\eqno(3.26)$$
But due to the second derivatives it is not a true equation of motion as
the $3\sigma$-equations. The adjoint equation reads
$$i\sigma^\nu_{a\bar a}\d_\nu\bar f^{\mu\bar a}=m f^\mu_{a}+
{2\over m}\d^\mu\d_\nu f^\nu_{a}.\eqno(3.27)$$

The next step is the decoupling of the fields. For this purpose we set
$$f_a^\mu=g_a^\mu+\alpha\d^\mu\lambda_a+\beta\sigma^\mu_{a\bar b}\bar
\lambda^{\bar b},\eqno(3.28)$$
$$\bar f_{\bar a}^\mu=\bar g_{\bar a}^\mu+\alpha^*\d^\mu\bar\lambda_ 
{\bar a}+\beta^*\sigma^\mu_{b\bar a}\lambda^b,\eqno(3.29)$$
and we choose the constants $\alpha, \beta$ in such a way that the $g$'s
satisfy the homogeneous $3\sigma$-equations
$$-i(\hat\sigma^\ro\sigma^\nu\hat\sigma^\mu)^{\bar a}_{\,c} 
\d_\nu\bar g_{\mu\bar a}+2m g^\ro_c=0,\eqno(3.30)$$
$$-i(\sigma^\ro\hat\sigma^\nu\sigma^\mu)_{a\bar c} 
\d_\nu g_\mu^a+2m\bar g^\ro_{\bar c}=0.\eqno(3.31)$$
This implies the relation
$$\alpha={i\over m}(1+2\beta^*).\eqno(3.32)$$
On the other hand, if we substitute the same form (3.28-29) into (3.26)
we see that all $\lambda$-terms cancel provided $\beta$ is real
$$\beta=\beta^ *.\eqno(3.33)$$

We remind the reader that all our fields are free quantum fields.
Therefore, the spinor field $\lambda(x)$ satisfies the anticommutation
relations (2.13). The situation is not so clear for the vector fields
because there exist different possibilities. Calculating the commutator
of (3.2) with $v_\mu+iw_\mu$, we find
$$[(v_\mu+iw_\mu)(x),(v_\nu+iw_\nu)(x')]=0,\eqno(3.34)$$
so that
$$[v_\mu(x),v_\nu(x')]=[w_\mu(x),w_\nu(x')]\=d C_{\mu\nu}(x-x'),\eqno(3.35)$$
assuming that all different fields commute. Similarly, computing the
commutators $[v_\mu(x), \{\bar Q_{\bar a},$ $\lambda_b(x')\}]$ and
$[v_\mu(x), \{Q_a, \lambda_b(x')\}]$, we obtain the relations
$$\{\bar f_{\mu\bar a}(x),\lambda_b(x')\}=-m\sigma^\alpha_{b\bar a}C_
{\mu\alpha}(x-x'),\eqno(3.36)$$
$$\{f_{\mu a}(x),\lambda_b(x')\}=-i(\sigma^\alpha\hat\sigma^\beta) 
_{ba}\d_\alpha C_{\mu\beta}(x-x').\eqno(3.37)$$
Substituting (3.29) into (3.36) and assuming that $\bar g^\mu_{\bar a}$
anticommutes with $\lambda_b$ we find
$$\alpha^*\sigma^\nu_{b\bar a}\d_\nu\d_\mu D(x-x')+im\beta^*
\sigma_{\mu b\bar a} D(x-x')=-m
\sigma^\nu_{b\bar a}C_{\mu\nu}(x-x').$$
Similarly (3.37) implies
$$im\alpha\eps_{ab}\d^\mu D-\beta (\sigma^\nu\hat\sigma^\mu)_{ba} 
\d_\nu D=-i(\sigma^\nu\hat\sigma^\beta)_{ba}\d_\nu C^{\mu}_\beta.\eqno(3.38)$$
We postpone the choice of the coefficients $\alpha, \beta$ to the next
section. 

Since we now have all commutators, we can write down the vector superfield
$V$ (3.1)
$$V=i\theta\lambda+i\theta\sigma^\nu\hat\sigma^\mu\theta\d_\nu(v_\mu+iw_\mu) 
-m\theta\sigma^\mu\bar\theta (v_\mu-iw_\mu)$$
$$+(\theta\sigma^\nu\hat\sigma^\mu\theta)(\bar\theta\d_\nu\bar f_\mu) 
-im(\theta\sigma^\mu\bar\theta)(\theta f_\mu)$$
$$-i{m\over 3}(\theta\sigma^\nu\hat\sigma^\mu\theta)(\bar\theta\bar 
\theta)\d_\nu (v_\mu-iw_\mu)+i{m\over 3}(\theta\sigma^\mu\bar\theta) 
(\theta\sigma^\nu\bar\theta)\d_\nu (v_\mu-iw_\mu)+{\rm h.c.}\eqno(3.39)$$
The last two terms can be rewritten in the form
$$i{m\over 2}(\theta\theta)(\bar\theta\bar 
\theta)\d^\mu (v_\mu-iw_\mu)+{\rm h.c.}$$
Furthermore, since $\theta\sigma^\nu\hat\sigma^\mu\theta=\theta\theta
\eta^{\nu\mu}$, we get the final result
$$V=i\theta\lambda+i(\theta\theta)\d^\mu(v_\mu+iw_\mu) 
-m\theta\sigma^\mu\bar\theta (v_\mu-iw_\mu)$$
$$+(\theta\theta)(\bar\theta\d^\mu\bar f_\mu) 
-im(\theta\sigma^\mu\bar\theta)(\theta f_\mu)$$
$$+i{m\over 2}(\theta\theta)(\bar\theta\bar 
\theta)\d^\mu (v_\mu-iw_\mu)+{\rm h.c.}\eqno(3.40)$$

We want to consider this field $V$ as a gauge superfield.
If we add the scalar field $S_2$ (2.39) to $V$, then the component fields are
changed consistently according to
$$m(v_\mu+iw_\mu)\to m(v_\mu+iw_\mu-{1\over m}\d_\mu F)\eqno(3.41)$$
$$m\d^\mu(v_\mu+iw_\mu)\to m\d^\mu(v_\mu+iw_\mu)-\sq F= 
m\d^\mu(v_\mu+iw_\mu)+m^2 F$$
$$\lambda\to\lambda+\chi\eqno(3.42)$$
$$f_{\mu a}\to f_{\mu a}+{i\over m}\d_\mu\chi_a.\eqno(3.43)$$
in all orders. The transformation (3.39) is the usual classical gauge
transformation. Therefore, the field $V$ is a good starting point for
supergauge theory. 
We will further investigate the properties of the vector superfield 
$V$ in the next section by analysing the mixed field $g^\mu_a(x)$. 
The free quantum superfields are the basis
of supergauge theory if considered as operator theory [6] in the 
spirit of a recent monograph [7].

\section{Investigation of the mixed field $g^\mu_a$}

The field $g$ must satisfy the homogeneous $3\sigma$-equations
(3.30-31) as well as the modified Dirac equations (3.26-27)
$$-i\sigma^\nu_{a\bar a}\d_\nu g^{\mu a}=m\bar g^\mu_{\bar a}+
{2\over m}\d^\mu\d_\nu\bar g^\nu_{\bar a},\eqno(4.1)$$
$$i\sigma^\nu_{a\bar a}\d_\nu\bar g^{\mu\bar a}=m g^\mu_{a}+
{2\over m}\d^\mu\d_\nu g^\nu_{a}.\eqno(4.2)$$
In addition we get constraints from
the anticommutators with the supercharges. The anticommutator (3.15) implies
$$\{\bar Q_{\bar a},g^\mu_{b}\}=-i\sigma^\nu_{b\bar a}\d_\nu
(v^\mu-iw^\mu)+\alpha m\sigma^ \nu_{b\bar a}  
\d^\mu(v_\nu-iw_\nu)$$ 
$$+i\beta(\sigma^ \mu\hat\sigma^ \alpha\sigma^ \beta)_{b\bar a} 
\d_\alpha(v_\beta-iw_\beta)\Bigl].\eqno(4.3)$$ 
and from (3.22) we obtain
$$\{Q_a,g^\mu_b\}=\eps_{ab}\Bigl[m(v^ \mu+iw^ \mu)+{2\over m}
\d^ \mu\d^\nu(v_\nu+iw_\nu)\Bigl]$$
$$+i\alpha(\sigma^\alpha\hat\sigma^\beta)_{ba}\d_\alpha\d^ \mu
(v_\beta+iw_\beta)-\beta m(\sigma^ \mu\hat\sigma^ \nu)_{ba}
(v_\nu+iw_\nu).\eqno(4.4)$$
These relations show that $g$ must be different from zero.

Since the vector index $\mu$ in $g^\mu_a$ corresponds to the spinor
representation $(\eh,\eh)$, the mixed field realizes the representation 
$(\eh,0)\times (\eh,\eh)$ of the proper Lorentz group. By the
Clebsch-Gordan decomposition it splits into irreducible representations
as follows
$$(\eh,0)\times (\eh,\eh)=(0,\eh)+(1,\eh).\eqno(4.6)$$
The dimensions of the representations are $2\times 4=2+6$.
We calculate with the most general (reducible) tensor
and set
$$g^{\mu a}=\sigma^\mu_{b\bar c}g^{ab\bar c}.\eqno(4.7)$$ 
The l.h.s is the so-called Rarita-Schwinger representation [8] of the spinor 
on the r.h.s.  The adjoint field is given by
$$\bar g^{\mu\bar a}=\sigma^\mu_{c\bar b}\bar g^{\bar a\bar bc}.\eqno(4.8)$$ 

If we substitute this representation into the $3\sigma$-equation (3.30)
we find the following Dirac equation for higher spin fields [8]
$$i\sigma^\nu_{a\bar c}\d_\nu\bar g^{\bar b\bar c}_{\;\;\; c}=m g_{ca} 
^{\;\;\;\bar b}.\eqno(4.9)$$
Similarly, the other $3\sigma$-equation (3.31) gives the adjoint Dirac
equation
$$i\hat\sigma^{\nu\bar ab}\d_\nu g_{cb}^{\;\;\;\bar c}=m\bar g^{\bar c 
\bar a}_{\;\;\;c}.\eqno(4.10)$$
The indices $c, \bar c$ are simply spectators in this Dirac equation.

Next we have to fulfill the equation (4.1). It
boils down to the second order equation
$$-i\sigma^\nu_{a\bar a}\d_\nu g^a_{\;\;d\bar d}=m\bar g_{\bar a\bar dd}
+{1\over m}\sigma^\mu_{d\bar d}\sigma^\nu_{c\bar b}\d_\mu\d_\nu 
\bar g_{\bar a}^{\;\;\bar bc}.\eqno(4.11)$$
Here we insert (4.9) in the form
$$\sigma^\nu_{c\bar b}\d_\nu\bar g^{\;\;\bar bc}_{\bar a}=
-im g^c_{\;\;c\bar a},\eqno(4.12)$$
and get
$$-i\sigma^\nu_{a\bar a}\d_\nu g^a_{\;\;d\bar d}
=m\bar g_{\bar a\bar dd}-i\sigma^\mu_{d\bar d}\d_\mu g^c_{\;\;c\bar a}. 
\eqno(4.13)$$
Now we decompose $g$ into a symmetric and antisymmetric part
$$g_{ad\bar d}=\eh(g_{ad\bar d}+g_{da\bar d})+\eh(g_{ad\bar d}-
g_{da\bar d})\eqno(4.14).$$
Since an antisymmetric second rank tensor has only one independent
component, we can write
$$g_{ad\bar d}=g'_{ad\bar d}+\eps_{ad}\bar\psi_{\bar d},\eqno(4.15)$$
$$\bar g_{\bar a\bar dd}=\bar g'_{\bar a\bar dd}+\eps_{\bar a\bar d}
\psi_d.$$
This is the decomposition (4.6) into irreducible Lorentz tensors.
As we will see, $\bar\psi$ is a true independent spinor field, and $g'$ 
is symmetric in the indices of the same kind.
We use this decompostion in the second Dirac equation (4.10) and obtain
$$i\sigma^\nu_{b\bar a}\d_\nu(g^{\prime cb}_{\;\;\;\bar c}+\eps^{cb} 
\bar\psi_{\bar c})=-m\bar g_{\bar c\bar a}^{\;\;\;c}.\eqno(4.16)$$
Substituting this into (4.13) we get
$$m(\bar g_{\bar a\bar d}^{\;\;\;d}-\bar g_{\bar d\bar a}^{\;\;\;d})=-2i
\sigma^{\nu d}_{\;\;\;\bar a}\d_\nu\bar\psi_{\bar d}+i
\sigma^{\nu d}_{\;\;\;\bar d}\d_\nu g^c_{\;\;c\bar a}.\eqno(4.17)$$ 

By (4.15) we can express $\bar\psi$ as a contracted tensor
$$2\bar\psi_{\bar d}=g^b_{\;\;b\bar d}.\eqno(4.18)$$
Then (4.17) gives the following equation for the spinor field $\psi$:
$$m\eps_{\bar a\bar d}\psi^ d=i(\sigma^{\nu d}_{\;\;\;\bar d}\d_\nu
\bar\psi_{\bar a}-\sigma^{\nu d}_{\;\;\;\bar a}\d_\nu\bar\psi
_{\bar d}).\eqno(4.19)$$
This is just the relation (2.40) which shows that
$\psi(x)$ is a spin-$\eh$ field satisfying the Dirac equation. Since all
steps can be inverted, the equation (4.11) is then satisfied. However,
the fields $g'$ and $\psi$ are not completely decoupled. In fact,
substituting (4.15) into (4.10) and using (4.19) we obtain an
inhomogeneous Dirac equation for $g'$:
$$i\hat\sigma_{\bar a}^{\nu b}\d_\nu g'_{bc\bar c}-m\bar g'_{\bar a 
\bar cc}=-i\sigma^\mu_{c\bar c}\d_\mu\bar\psi_{\bar a}.\eqno(4.20)$$
Consequently, $g'_{ab\bar c}(x)$, although being symmetric in $a, b$, is
not identical with the Rarita - Schwinger field for spin-3/2 particles [8].

Finally, there remains to check the consistency of the anticommutators 
with the supercharge, because this is the point where something goes
wrong if there is an incorrect step in the construction. The
anticommutator (4.4) gives the
corresponding anticommutator for the spinor field:
$$\{Q_a, g_{bc\bar c}\}=\eps_{ab}\sigma^\mu_{c\bar c}\Bigl[{m\over 2}
(v_\mu+iw_\mu)+\d_\mu\d^\nu(v_\nu+iw_\nu)({1\over m}+{i\over 2}\alpha)
\Bigl]$$
$$-\beta m\eps_{cb}\hat\sigma^\nu_{\bar c a}(v_\nu+iw_\nu)+\alpha 
\sigma^\mu_{c\bar c}\sigma^{\alpha\beta}_{ba}\d_\alpha\d_\mu (v_\beta 
+iw_\beta).\eqno(4.21)$$
For any choice of $\alpha$ and $\beta$ the r.h.s. can be decomposed into
a symmetric and antisymmetric part with respect to $b, c$. This allows
to identify the anticommutators of $g'$ and $\psi$ and shows the
consistency of the whole construction. A preferred choice is
$$\alpha=0,\quad \beta=-{1\over 2},\eqno(4.22)$$
because by (3.36) this leads to the propagator
$$C_{\mu\nu}(x-x')={i\over 2}\eta_{\mu\nu}D(x-x')\eqno(4.23)$$
which has a good ultraviolet behaviour. For this choice let us write down  
the antisymmetric part of (4.21):
$$\{Q_a,\bar\psi_{\bar c}\}=-{1\over 2}\sigma^\mu_{a\bar c}\Bigl[{3\over
2}m(v_\mu+iw_\mu)+{1\over m}\d_\mu\d^\nu(v_\nu+iw_\nu)\Bigl].
\eqno(4.24)$$
Here we have used the simple identity
$$\eps_{ab}\sigma^\mu_{c\bar c}-\eps_{ac}\sigma^\mu_{b\bar c}=
\eps_{cb}\sigma^\mu_{a\bar c}.\eqno(4.25)$$

For later use we also calculate the anti-commutation relations of the
mixed fields. By means of (3.12) we find
$$\{f^\mu_a(x),\bar f^\nu_{\bar b}(x')\}=-\{f^\mu_a(x),[\bar Q_{\bar b}
, v^\nu(x')]\}=[v^\nu(x'),\{f^\mu_a(x),\bar Q_{\bar b}\}]$$
$$=i\sigma^\ro_{a\bar b}\d^x_\ro[v^\mu(x)-iw^\mu(x), v^\nu(x')]=$$
$$=i\sigma^\ro_{a\bar b}\d^x_\ro C^{\mu\nu}(x-x')=-{1\over 2}\eta^
{\mu\nu}\sigma^\ro_{a\bar b}\d^x_\ro D(x-x')\eqno(4.26)$$
for the preferred choice (4.23). This implies the following
anti-commutator for the mixed $g$-field:
$$\{g^\mu_a(x),\bar g^\nu_{\bar b}(x')\}=-{1\over 2}\eta^
{\mu\nu}\sigma^\ro_{a\bar b}\d^x_\ro D(x-x')-{1\over 4}(\sigma^\mu
\hat\sigma^\ro\sigma^\nu)_{a\bar b}\d_\ro D(x-x').\eqno(4.27)$$
In a similar way we obtain
$$\{f^\mu_a(x),f^\nu_b(x')\}=-{i\over 2}\eps_{ab}(m\eta^{\mu\nu}+
{2\over m}\d^\mu\d^\nu)D(x-x'),\eqno(4.28)$$
and
$$\{g^\mu_a(x),g^\nu_b(x')\}=-{i\over 4}m(\eps_{ab}\eta^{\mu\nu}- 
2i\sigma^{\mu\nu}_{ab})D(x-x')-{i\over m}\eps_{ab}\d^\mu\d^\nu 
D(x-x'),\eqno(4.29)$$
$$\{\bar g^\mu_{\bar a}(x),\bar g^\nu_{\bar b}(x')\}={i\over 4}m 
(\eps_{\bar a\bar b}\eta^{\mu\nu}+2i\hat\sigma^{\mu\nu}_{\bar a\bar b} 
)D(x-x')+{i\over m}\eps_{\bar a\bar b}\d^\mu\d^\nu 
D(x-x'),\eqno(4.30)$$
$$\{\bar g^\mu_{\bar a}(x),g^\nu_b(x')\}=-{1\over 2}\eta^
{\mu\nu}\sigma^\ro_{b\bar a}\d^x_\ro D(x-x')-{1\over 4}(\sigma^\nu
\hat\sigma^\ro\sigma^\mu)_{b\bar a}\d_\ro D(x-x').\eqno(4.31)$$

It is interesting to discuss our results from the point of view of
representation theory. The massive representations of the $N=1$ 
supersymmetric algebra have the following spin components:

\vskip 0.5cm
{\offinterlineskip \tabskip=0pt
\halign{ \strut \vrule#&\quad \hfil #\quad &\vrule#&\quad \hfil #\quad &
\vrule#&\quad \hfil #\quad &\vrule#& \quad \hfil #\quad &\vrule#& 
\quad \hfil #\quad &\vrule#\cr
\noalign{\hrule} 
&{\rm spin} && $\Omega_0$ && $\Omega_{1/2}$ && $\Omega_1$ && $\Omega_{3/2}$
&\cr \noalign{\hrule} 
& 0 && 2 && 1 && && & \cr
& 1/2 && 1 && 2 && 1 &&  & \cr
& 1 && && 1 && 2 && 1 & \cr
& 3/2 && && && 1 && 2 & \cr
& 2 && && && && 1 & \cr
\noalign{\hrule}}}
\vskip 0.5cm
\noindent The representation $\Omega_0$ corresponds to the scalar 
superfield $S_1$
(2.33) because it has 2 scalar and 1 spinor component. The vector
superfield $V$ (3.40) may be related to $\Omega_1$, if we consider the
mixed field $f^\mu_a$ as the spin-3/2 component. But this is not
entirely justified because $f^\mu_a$ contains also a spin-1/2 piece. The
surprising result is that it was impossible to realize the
representation $\Omega_{1/2}$ with free quantum fields. In fact, if we
start with the single scalar component required by this representation,
as we did in sect.2, we are driven into the representation $\Omega_0$.

\end{document}